\documentclass[superscriptaddress,preprintnumbers,amsmath,amssymb]{revtex4}
\usepackage[dvips]{graphicx}

\def\ltap{\ \raise.3ex\hbox{$<$\kern-.75em\lower1ex\hbox{$\sim$}}\ }
\def\gtap{\ \raise.3ex\hbox{$>$\kern-.75em\lower1ex\hbox{$\sim$}}\ }
\def\gl{\ \raise.4ex\hbox{$>$\kern-.75em\lower1ex\hbox{$<$}}\ }

\begin{document}

\vspace*{-1.5cm}
\begin{flushright}
KEK-TH-1349
\end{flushright}

\title{Fermion structure of non-Abelian vortices in high density QCD}
\author{Shigehiro Yasui}
\affiliation{KEK Theory Center, Institute of Particle 
and Nuclear Studies,
High Energy Accelerator Research Organization (KEK),
1-1 Oho, Tsukuba, Ibaraki, 305-0801, Japan}

\author{Kazunori Itakura} 
\affiliation{KEK Theory Center, Institute of Particle 
and Nuclear Studies,
High Energy Accelerator Research Organization (KEK),
1-1 Oho, Tsukuba, Ibaraki, 305-0801, Japan}
\affiliation{Department of Particle and Nuclear Studies, 
Graduate University for Advanced Studies (SOKENDAI), 
1-1 Oho, Tsukuba, Ibaraki, 305-0801, Japan}

\author{Muneto Nitta}
\affiliation{Department of Physics, and Research and Education Center 
for Natural Sciences, Keio University, 4-1-1 Hiyoshi, Yokohama, 
Kanagawa 223-8521, Japan}

\date{January 2010}

\begin{abstract}
We study the internal structure of a non-Abelian 
vortex in color superconductivity with respect to quark
degrees of freedom. Stable non-Abelian vortices 
appear in the Color-Flavor-Locked phase whose symmetry 
$SU(3)_{\rm c+L+R}$ is further broken to 
$SU(2)_{\rm c+L+R}\ \otimes$ $U(1)_{\rm c+L+R}$ at the vortex 
cores. Microscopic structure of vortices at scales 
shorter than the coherence length can be analyzed by the 
Bogoliubov-de Gennes (B-dG) equation (rather than the 
Ginzburg-Landau equation). We obtain quark spectra from 
the B-dG equation by treating the diquark gap having the 
vortex configuration as a background field. We find that 
there are massless modes (zero modes) well-localized around 
a vortex, in the triplet and singlet states of the 
unbroken symmetry $SU(2)_{\rm c+L+R}\ 
\otimes$ $U(1)_{\rm c+L+R}$. 
The velocities $v_i$ of the massless modes ($i=t,s$ for triplet 
and singlet) change at finite chemical potential $\mu\neq 0$, 
and decrease as $\mu$ becomes large. Therefore, low energy 
excitations in the vicinity of the vortices are effectively 
described by 1+1 dimensional massless fermions whose 
velocities are reduced $v_i<1$. 
\end{abstract}

\maketitle

\section{Introduction}

The structure of vortices is crucial in understanding
the dynamics of the phases in a broad subject of physics from 
condensed-matter physics to high-energy physics and 
astrophysics. Macroscopic dynamics of vortices is described 
by a spatially dependent order parameter for a certain 
broken symmetry. It is well described by the Ginzburg-Landau 
theory,  which has conducted great 
progress in studies of vortices in superconductors 
at the length scales longer 
than the coherence length and the penetration depth. On the 
other hand, in order to understand short range structure of 
vortices such as the vortex core, one needs a microscopic 
theory in which vortices are described by the fermionic 
degrees of freedom. 
Such a self-consistent description including both the order 
parameter and fermions is known as the Bogoliubov-de Gennes 
(B-dG) equation \cite{deGennes}. The first investigation of 
the vortex structure by solving the B-dG equation was done 
in Ref.~\cite{Caroli:1964}. Fermions are trapped inside the 
vortex core if their energies are less than the gap of the 
ground state while they are scattering states if their energies 
are larger than it. Since this first study treated the order 
parameter (the superconducting gap) as a background field (and thus 
it is not self-consistent), successive researches were 
devoted towards finding the self-consistent solutions to 
the B-dG equation \cite{Shore:1989,Gygi:1990}. The complete 
self-consistent description was achieved by considering 
both quasi-bound and scattering fermions in the vortex 
\cite{Gygi:1991zz,Machida:2003}. These studies predicted 
an enhanced local density of fermion states at the Fermi 
level around the vortex core, 
which is experimentally observed in various materials 
\cite{Sacepe:2006,Guillamon:2008}.
Recently the analysis of vortices in the B-dG equation 
has been also applied to the BEC-BCS crossover in 
fermionic cold atom systems 
\cite{Machida:2005,Sensarma:2006}. 

Among all fermion modes associated with vortices,
zero modes (gapless fermions) play an especially 
important role in various subjects of physics. 
In non-relativistic theories 
including all condensed matter systems, 
fermion ``zero modes" trapped in a vortex core naturally appear
in the semi-classical approximation, but they are actually not
necessarily exactly zero-energy states because they in general
acquire small gaps $\sim \Delta^2/E_{\rm F}$ 
(called ``minigaps") that are small enough
compared with the gap $\Delta$ \cite{Volovik}.
For instance, a conventional $s$-wave superconductor is 
such a case \cite{Caroli:1964}. 
In chiral $p$-wave superconductors Majorana fermions 
trapped inside a vortex core have exactly zero energy, which
attain non-Abelian statistics \cite{Read:2000,Ivanov:2001} 
and give a candidate of quantum computations.
Recently it has been found that even a 
conventional $s$-wave superconductor 
allows fermion zero modes in a vortex core when it is 
surrounded by a topological insulator \cite{Fu:2008zzb}.
Unlike non-relativistic theories, 
fermion zero modes in relativistic theories 
\cite{Jackiw:1981ee} have exactly zero energy, 
and 
the number of zero modes is determined by the index theorem 
to be $2n$ for $n$ winding vortices \cite{Weinberg:1981eu}.
Fermion zero modes are also important in the context of
cosmic strings 
\cite{Witten:1984eb}.

The study of vortices has been extended to color superconductivity 
in dense quark matter, where the attractive force between quarks 
induces a rich structure of symmetry breaking patterns 
\cite{Alford:2007xm}. At extremely high densities, a novel 
phase called the color-flavor-locked (CFL) phase will be 
realized, where up, down and strange quarks participate in 
the Cooper pairing with symmetry breaking pattern 
$SU(3)_{\rm c} \otimes SU(3)_{\rm L} 
\otimes SU(3)_{\rm R} \rightarrow SU(3)_{\rm c+L+R}$ 
\cite{Alford:1998mk,Alford:1999pa,Alford:2007xm}.
The color-flavor structure of the gap in the CFL phase 
is given by $\Delta^{\alpha i} 
\propto \epsilon^{\alpha \beta \gamma} \epsilon^{ijk}  
\langle \psi^{\beta j} \psi^{\gamma k} \rangle$, where
$i$, $j$, $k$ are flavor indices, and $\alpha$, $\beta$,
$\gamma$ are color indices. It was argued that 
superfluid vortices would exist as a result of breaking 
of the global $U(1)_{\rm B}$ symmetry of the baryon number 
\cite{Forbes:2001gj,Iida:2002ev,Iida:2004if}.
Color flux tubes have been suggested 
\cite{Iida:2002ev,Iida:2004if} but they are 
topologically and dynamically unstable. 
The true ground states of vortices in the CFL phase are 
non-Abelian vortices found by Balachandran, Digal and 
Matsuura \cite{Balachandran:2005ev} 
which are topologically stable superfluid vortices 
carrying color magnetic fluxes inside their core. 
It is a 1/3 quantized vortex, 
namely it has 1/3 $U(1)_{\rm B}$ winding (circulation) 
and 1/9 tension of 
those of a global $U(1)_{\rm B}$ vortex 
\cite{Forbes:2001gj,Iida:2002ev,Iida:2004if}, 
and it also has 1/3 amount of the color flux of 
that of a topologically and dynamically
unstable color flux tube \cite{Iida:2002ev,Iida:2004if}. 
The detailed gap profile function and the size of color 
flux has been calculated in Ref.~\cite{Eto:2009kg} 
in the Ginzburg-Landau approach. 
One characteristic property of non-Abelian vortices is that  
the color-flavor-locked symmetry $SU(3)_{\rm c+L+R}$ 
of the ground state CFL phase is spontaneously broken down to 
its subgroup $SU(2)_{\rm c+L+R} \otimes U(1)_{\rm c+L+R}$ 
in the presence of a non-Abelian vortex. 
According to this symmetry breaking, Nambu-Goldstone 
zero modes appear around the vortex, which parametrize 
a coset space
$
\mathbf{C}P^{2} \simeq SU(3)_{\rm c+L+R} 
 / [SU(2)_{\rm c+L+R} \otimes U(1)_{\rm c+L+R}]
$ \cite{Nakano:2007dq,Nakano:2007dr}. Therefore, the vortex 
solution allows a continuous family of solutions with 
degenerate energy,  corresponding to $\mathbf{C}P^{2}$. 
These modes are called the internal orientational zero modes  
and each point of $\mathbf{C}P^{2}$ corresponds to the color flux 
which the non-Abelian vortex carries. 
Since this breaking occurs in the vicinities of the vortex, 
these modes are localized around it, and  
one can construct a 1+1 dimensional effective theory 
on the vortex world-sheet by integrating 
the original action over the vortex codimensions \cite{Eto:2009bh}.
The interaction between two non-Abelian vortices at large distance 
is mediated by fluctuations of $U(1)_{\rm B}$ Nambu-Goldstone modes, 
yielding the universal repulsion between them 
irrespective to their $\mathbf{C}P^{2}$ orientations in the 
internal space \cite{Nakano:2007dr,Nakano:2008dc}. 
This predicts that each global $U(1)_{\rm B}$ vortex is 
dynamically unstable 
to decay into three non-Abelian vortices which 
have different color fluxes with the total flux cancelled out. 
Furthermore, it also predicts a non-Abelian vortex lattice 
at least when the lattice spacing is much larger than the vortex core size
\cite{Nakano:2007dr}. Such a vortex lattice is expected to 
be realized in a CFL quark matter core of a rapidly rotating 
neutron (or compact) star
\cite{Sedrakian:2008ay,Shahabasyan:2009zz}.

However, these studies are all based on a macroscopic theory, 
namely the Ginzburg-Landau model which is valid only at the 
scale larger than the penetration depth and coherence length.
In order to understand the whole structure of 
the non-Abelian vortices including the region inside their cores, 
we have to consider fermion dynamics from the B-dG equation.
The most important fermion modes are the zero modes as mentioned 
which exist exactly at the Fermi energy. 
The purpose of the present paper is to study the fermion zero modes 
in a non-Abelian vortex, based on the B-dG equation.

This paper is organized as follows. 
In Sec.~\ref{Sect:B-dG} we construct the fermion zero modes 
around a vortex background in the cases of a single flavor 
and the CFL phase. 
We find four fermion zero modes for a non-Abelian vortex 
in the CFL phase, which belong to the singlet and triplet 
representations of the unbroken symmetry $SU(2)_{\rm c+L+R}$. 
We find an interesting layer structure of the zero modes in 
the core of the vortices; the core size of the singlet 
zero mode is the half of the one of the triplet zero modes.
In Sec.~\ref{Sect:EFT} we construct the 1+1 dimensional 
effective field theory of the fermion zero modes. 
We find that the velocity of the zero modes propagating 
along the vortex line depend on the chemical potential, 
and that the velocity of singlet is about quarter of that of 
the triplet at high baryon density. 
The last section is devoted to summary of the results and
discussion of the future problems.

\section{Bogoliubov-de Gennes equation in color superconductivity}
\label{Sect:B-dG}
\subsection{Single flavor}
As the simplest example which shares the essential physics
with the CFL case,  we first discuss a relativistic 
fermion system with a single flavor that exhibits superconductivity 
and allows for Abelian vortices. We treat a single (or, an isolated) 
vortex whose core is located on an infinite straight line along the $z$ axis,
and assume that the winding number (vorticity) is one
\footnote{Since the energy of a single vortex $\Delta=|\Delta| 
\, {\rm e}^{i n \theta}$ with winding number 
$|n|\ge2$ is proportional to $n^2$, it decays to $n$ vortices 
each of which has a unit vorticity. Therefore, we consider only
the case with $n=1$. Note also that the anti-vortex with $n=-1$ 
gives qualitatively the same results.}.
As long as we treat the gap profile
function $\Delta(r)$ as a background field, the description in 
the present subsection is very similar to that of the 
vortex-fermion system discussed in 
Refs.~\cite{Jackiw:1981ee,Weinberg:1981eu,Witten:1984eb}.
However, we consider here a fermionic matter at {\it finite 
densities} (otherwise color superconductivity does not take place) 
while the previous studies 
\cite{Jackiw:1981ee,Weinberg:1981eu,Witten:1984eb} 
were formulated only in the vacuum. 
As we will see later, the effects of finite densities are 
quite important for low energy excitations of the fermion
modes in a vortex. 

The Bogoliubov-de Gennes (B-dG) equation is useful when one 
considers inhomogeneous superconductivity \cite{deGennes}: 
in the presence of normal-super 
boundaries and impurities. Among such examples is the vortex 
in which the gap is locally reduced compared to the value in the 
homogeneous ground state. While the Ginzburg-Landau equation 
is usually used to find a vortex profile, it is valid only for 
spatial variations at scales longer than the coherent length. 
On the other hand, since the B-dG equation is represented by 
microscopic (fermion) degrees of freedom, it can in principle 
describe the internal structure of a vortex at shorter length 
scales. For the present case, it is expressed in the Nambu-Gor'kov
(particle-hole) representation as  
\begin{eqnarray}
{\cal H}\Psi = {\cal E}\Psi\, ,
\label{B-dG}
\end{eqnarray}
with the Hamiltonian density in the mean-field approximation 
\begin{eqnarray}
{\cal H}=\left(
\begin{array}{cc}
 -i\gamma_{0} \vec{\gamma} \cdot \vec{\nabla}-\mu  & \Delta(x) \gamma_{0} \gamma_{5}     \\
 -\Delta^{\ast}(x) \gamma_{0} \gamma_{5} & -i\gamma_{0} \vec{\gamma} \cdot \vec{\nabla}+\mu
\end{array}
\right)\, .
\label{eq:hamil_single}
\end{eqnarray}
Here, we consider a relativistic massless fermion 
and the energy is evaluated with respect to the Fermi surface
($\mu$ is the chemical potential of a fermion).  
The gap is supposed to take a vortex profile : 
 $\Delta(x)=|\Delta(r)|\, {\rm e}^{i\theta}$ 
with $|\Delta(r=0)|= 0$ and $|\Delta(r=\infty)|=|\Delta|$
(where $r=\sqrt{x^2+y^2}$ and $\theta$ is the angle 
around the vortex).

As the first attempt to this problem, we treat the gap 
function as a background field as was done in Ref.~\cite{Caroli:1964}.
However, strictly speaking, the B-dG equation (\ref{B-dG}) should be 
solved self-consistently because the gap profile function is 
expressed in terms of fermion degrees of freedom. Such a
self-consistent description has been discussed in a simpler 
configuration like a kink, and recently it was achieved even 
for vortices in a standard type-II superconductor 
\cite{Gygi:1991zz}. We leave such a description in the color 
superconductivity for future problems.

With the cylindrical shape of the vortex profile, one can 
write the eigenstates of Eq.~(\ref{B-dG}) as 
$\Psi_{\pm,m}^{k_z}(r,\theta,z)$ with 
\begin{eqnarray}
\Psi_{\pm,m}^{k_{z}}(r,\theta,z)
=\Psi_{\pm,m}(r,\theta)\, {\rm e}^{i k_{z} z},
\end{eqnarray}
which has the momentum $k_{z}$ in the $z$ direction, and 
chirality $\pm$ corresponding to right and left.
The Nambu-Gor'kov structure of $\Psi_{\pm,m}(r,\theta)$ is 
expressed as 
\begin{eqnarray}
\Psi_{\pm,m}(r,\theta)=
\left(
\begin{array}{c}
 \varphi_{\pm,m}(r,\theta)  \\
 \eta_{\mp,m-1}(r,\theta)
\end{array}
\right),
\end{eqnarray}
where particle $\varphi_{\pm,m}$ and hole $\eta_{\mp,m-1}$
components are eigenstates of the $z$ component of the total 
spin operator $J_{z}$ with their eigenvalues $m+1/2$ and 
$(m-1)+1/2$, respectively \cite{Jackiw:1981ee}. 
We note that the chiralities of a particle and a hole (defined 
by the eigenvalues of $\gamma_{5}$) are 
$\pm$ and $\mp$ for right and left modes, respectively.

Once the concrete form of the vortex profile $|\Delta(r)|$ 
is given, one can explicitly solve the B-dG equation (\ref{B-dG}). 
However, we are rather interested in low energy excitations 
which may be independent of the precise form of the profile. 
In fact, even without knowing the precise form of the profile, 
one can address the questions whether there exist zero modes or 
not, and if they exist, what are the properties of the zero modes. 
Note that these are highly nontrivial questions: in a homogeneous 
superconductor, even though the original fermion is massless, 
there is no excitation around the Fermi surface 
below the gap energy. Of course, the B-dG equation reduces to 
the standard Bogoliubov equation for a spatially constant gap, 
and there is no massless excitation in that case. On the other hand, 
since it is known that the vortex-fermion system {\it in the vacuum} 
allows for zero modes which are trapped in a vortex 
\cite{Jackiw:1981ee,Witten:1984eb}, and that its existence is 
guaranteed from the topological considerations \cite{Weinberg:1981eu}, 
one naturally expects that there exist zero-mode solutions 
{\it even at finite densities}.  
This is indeed the case. The B-dG equation at finite densities 
(\ref{B-dG}) for a generic profile function $|\Delta(r)|$ 
allows for zero-mode solutions, namely, solutions with the 
${\cal E}=0$ eigenvalue. The zero modes are realized for 
$m=0$ and $k_z=0$, 
and the explicit forms can be written as, for the right mode 
($+$ for a particle, $-$ for a hole)  
\begin{eqnarray}
\varphi_{+,0}(r,\theta) &=& 
C \, {\rm e}^{-\int_{0}^{r}|\Delta(r')|\, \mbox{d}r'}
\left(
\begin{array}{c}
 J_{0}(\mu r) \\
 i J_{1}(\mu r)\, {\rm e}^{i \theta}
\end{array}
\right), \label{one_flavor_left_particle}\\
\eta_{-,-1}(r,\theta) &=& 
C\, {\rm e}^{-\int_{0}^{r}|\Delta(r')|\, \mbox{d}r'}
\left(
\begin{array}{c}
 -J_{1}(\mu r) \, {\rm e}^{-i \theta}\\
 i J_{0}(\mu r) 
\end{array}
\right),\label{one_flavor_left_hole}
\end{eqnarray}
and for the left mode ($-$ for a particle, $+$ for a hole)  
\begin{eqnarray}
\varphi_{-,0}(r,\theta) &=& 
C^{\prime}\, {\rm e}^{-\int_{0}^{r}|\Delta(r')|\, \mbox{d}r'}
\left(
\begin{array}{c}
 J_{0}(\mu r) \\
- i J_{1}(\mu r)\, {\rm e}^{i \theta}
\end{array}
\right), \label{one_flavor_right_particle}\\
\eta_{+,-1}(r,\theta) &=& 
C^{\prime} \, {\rm e}^{-\int_{0}^{r}|\Delta(r')|\, \mbox{d}r'}
\left(
\begin{array}{c}
 J_{1}(\mu r) \, {\rm e}^{-i \theta}\\
 i J_{0}(\mu r) 
\end{array}
\right),\label{one_flavor_right_hole}
\end{eqnarray}
where $C$ and $C^{\prime}$ are normalization constants, 
and $J_n(x)$ is the Bessel function. 
We have represented the solutions in the Weyl 
(2-component) spinors. 
In obtaining the solutions above, 
we have imposed finiteness at $r=0$ and in the limit $r\to \infty$. 
When $\mu= 0$, these solutions recover the zero-mode solutions 
in the vacuum found in Ref.~\cite{Jackiw:1981ee}, as they should. 
Notice that the zero-mode solutions above satisfy the following 
``Majorana-like" condition \footnote{The Majorana 
condition with $\kappa=1$ leads to the right mode, 
while $\kappa=-1$ the left mode.} 
($\kappa=\pm 1$): 
\begin{eqnarray}
\Psi = \kappa \, {\cal U}\, \Psi^{\ast}, \hspace{1em}
{\cal U}=
\left(
\begin{array}{cc}
 0 & \gamma_{2} \\
 \gamma_{2} & 0   
\end{array}
\right)\, ,
\label{eq:Majorana}
\end{eqnarray}
which leads to ${\cal U}^{-1} {\cal H U}=-{\cal H}^{\ast}$ 
and thus physically implies the equivalence between a particle 
and a hole. 
One can explicitly check that if the solution satisfies this 
condition, it indeed leads to $m=0$, $k_{z}=0$ and ${\cal E}=0$.

In the limit of small size of the vortex, in other words, 
at a distance much away from the vortex, $|\Delta(r)|$ can be 
regarded as a constant $|\Delta|$ which is equal to the gap 
in the bulk space.
Then, for example, the right mode is given by a simpler form
\begin{eqnarray}
\varphi_{+,0}(r,\theta) &\simeq & C \, {\rm e}^{-|\Delta|r}
\left(
\begin{array}{c}
 J_{0}(\mu r) \\
 i J_{1}(\mu r)\, {\rm e}^{i \theta}
\end{array}
\right), \\
\eta_{-,-1}(r,\theta) &\simeq& C \, {\rm e}^{-|\Delta|r}
\left(
\begin{array}{c}
 -J_{1}(\mu r)\, {\rm e}^{-i \theta} \\
 i  J_{0}(\mu r) 
\end{array}
\right).
\end{eqnarray}
This asymptotic behavior clearly shows that the zero-mode solutions 
are well-localized around the vortex line.

\subsection{Color-Flavor-Locked phase}
\subsubsection{B-dG equation with a non-Abelian vortex}
Now we discuss the fermion structure of the non-Abelian vortex
which appears in the CFL phase. 
The gap configuration of the non-Abelian vortex in the color-flavor
representation $\Delta^{\alpha i}
\propto \epsilon^{\alpha\beta\gamma}\epsilon^{ijk}
\langle \psi^{\beta j {\rm T}} C\gamma_5 \psi^{\gamma k}\rangle $
($C=i\gamma^2\gamma^0$ is the charge conjugation, 
$\alpha,\beta,\gamma$ and $i,j,k$ are the 
color and flavor indices, respectively)
is given by \cite{Balachandran:2005ev} 
\begin{eqnarray}
\Delta(r,\theta) =
\left(
\begin{array}{ccc}
 \Delta_{1}(r,\theta) & 0  & 0  \\
 0 & \Delta_{0}(r)  & 0  \\
 0 & 0 & \Delta_{0}(r)  
\end{array}
\right)\, , \label{gap_non-Abelian}
\end{eqnarray}
where $\Delta_{1}(r,\theta) = |\Delta_{1}(r)|\, {\rm e}^{i\theta}$ 
corresponds to the vortex configuration with winding number one, 
and $\Delta_{0}(r)$ does not have a winding number (though it is 
not spatially constant). 
If two gaps are the same and constant $\Delta_1=\Delta_0
=\Delta_{\rm CFL}$, it is the CFL phase which is symmetric 
under the rotation in $SU(3)_{\rm c+L+R}$. 
On the other hand, the gap structure (\ref{gap_non-Abelian})
is invariant only under the transformation in 
$SU(2)_{\rm c+L+R}\, \otimes\, U(1)_{\rm c+L+R}$
which is a subgroup of $SU(3)_{\rm c+L+R}$. 
Still, the symmetry of the bulk CFL phase is restored\footnote{Precisely,
the gap approaches $\Delta (r\to \infty) = 
\Delta_{\rm CFL} {\rm diag}. ({\rm e}^{i\theta}, 1,1) $, and the extra 
phase in the first component can be factored out by 
a regular gauge transformation as 
$\Delta \to U \Delta = \Delta_{\rm CFL}\, {\rm e}^{i\theta/3} 
{\rm diag}.(1,1,1)$. 
Here $U(r,\theta)$ is a {\it regular} gauge transformation 
given by 
$U(r,\theta) = \exp [i g(r) (\theta/3) {\rm diag.} (-2,1,1)]\in SU(3)$ 
where $g(r)$ is an arbitrary function 
with the boundary conditions $g(r=0) = 0$ and 
$g(r\to \infty) = 1$ \cite{Nakano:2007dr}.
} 
in the limit $r\to \infty$ because the gap profile 
function $|\Delta_1(r)|$ approaches the value of the CFL 
phase $|\Delta_1(r)|\to |\Delta_{\rm CFL}|$. 

For the problem of our interest, it is sufficient to 
consider the following mean-field Hamiltonian density which 
exhibits the CFL phase:
\begin{equation}
{\cal H}= 
\psi_i^{\alpha\dagger} 
(-i\gamma_0 \vec{\gamma}\cdot \vec{\nabla}-\mu)
\psi_i^\alpha  
+ \left[
\Delta_{ij}^{\alpha\beta} 
(\psi^{\alpha {\rm T}}_iC\gamma_5 \psi^{\beta}_j )^\dagger 
+ \, {\rm h.c.}
\right]
\, ,
\end{equation}
where the gap is now expressed as 
$\Delta^{\alpha\beta}_{ij}\propto 
\langle \psi^{\alpha{\rm T}}_i C\gamma_5 \psi^{\beta}_j\rangle $. 

An explicit representation of the B-dG equation 
${\cal H}\Psi = {\cal E}\Psi$ is given by (for a similar 
representation in the homogeneous CFL phase, see 
Refs.~\cite{Alford:1999pa,Sadzikowski:2002in})
\begin{eqnarray}
\left(
\begin{array}{ccccccccc}
 \hat{\cal H}_0 & \hat{\Delta}_{1} & \hat{\Delta}_{0} & 0 & 0 & 0 & 0 & 0 & 0 \\
 \hat{\Delta}_{1} & \hat{\cal H}_0 & \hat{\Delta}_{0} & 0 & 0 & 0 & 0 & 0 & 0 \\
 \hat{\Delta}_{0} & \hat{\Delta}_{0} & \hat{\cal H}_0 & 0 & 0 & 0 & 0 & 0 & 0 \\
 0 & 0 & 0 & \hat{\cal H}_0 & -\hat{\Delta}_{1} & 0 & 0 & 0 & 0 \\
 0 & 0 & 0 & -\hat{\Delta}_{1} & \hat{\cal H}_0  & 0 & 0 & 0 & 0 \\
 0 & 0 & 0 & 0 & 0 & \hat{\cal H}_0 & -\hat{\Delta}_{0} & 0 & 0 \\
 0 & 0 & 0 & 0 & 0 & -\hat{\Delta}_{0} & \hat{\cal H}_0 & 0 & 0 \\
 0 & 0 & 0 & 0 & 0 & 0 & 0 & \hat{\cal H}_0 & -\hat{\Delta}_{0} \\
 0 & 0 & 0 & 0 & 0 & 0 & 0 & -\hat{\Delta}_{0} & \hat{\cal H}_0
\end{array}
\right)
\left(
\begin{array}{c}
 u_r \\
 d_g \\
 s_b \\
 d_r \\
 u_g \\
 s_r \\
 u_b \\
 s_g \\
 d_b
\end{array}
\right)
= {\cal E}
\left(
\begin{array}{c}
 u_r \\
 d_g \\
 s_b \\
 d_r \\
 u_g \\
 s_r \\
 u_b \\
 s_g \\
 d_b
\end{array}
\right),
\label{eq:CFL_eve} 
\end{eqnarray}
where we have introduced the notation e.g., 
 $u_{r}$ for  the quark with color ``red" and flavor ``up" in
 the Nambu-Gor'kov representation.
The matrices $\hat{\cal H}_{0}$ and $\hat{\Delta}_{i}$ 
($i=0$ and 1) are given as follows: 
\begin{eqnarray}
\hat{\cal H}_{0} &=&
\left(
\begin{array}{cc}
 -i\gamma_{0} \vec{\gamma} \!\cdot\! \vec{\nabla} - \mu & 0 \\
 0 & -i\gamma_{0} \vec{\gamma} \!\cdot\! \vec{\nabla} + \mu
\end{array}
\right), \\
\hat{\Delta}_{i} &=&
\left(
\begin{array}{cc}
 0 & \Delta_{i} \gamma_{0} \gamma_{5} \\
 -\Delta_{i}^{\dag} \gamma_{0} \gamma_{5} & 0
\end{array}
\right).
\end{eqnarray} 

\subsubsection{Multiplets in $SU(2)_{\rm c+L+R}$ }

Eigenstates of Eq.~(\ref{eq:CFL_eve}) are classified into 
multiplets of the unbroken $SU(2)_{\rm c+L+R}$ symmetry:  
triplet, doublet and singlet states. Note first that 
nine quark states (3 flavors $\times$ 3 colors) can be 
decomposed by the generators of the $SU(3)_{\rm c+L+R}$ 
symmetry of the CFL phase together with a unit matrix 
\cite{Sadzikowski:2002in}:
\begin{eqnarray}
\left(
\begin{array}{ccc}
 u_r & u_g  & u_b  \\
 d_r & d_g & d_b  \\
 s_r & s_g & s_b  
\end{array}
\right) =
\sum_{A=1}^{9}  \Psi^{(A)}\frac{\lambda_A}{\sqrt{2}}\, ,
\label{eq:CFL_base}
\end{eqnarray}
where $\lambda_A\, (A=1,\cdots,8)$ are the Gell-Mann 
matrices normalized as $\mbox{tr} (\lambda_A \lambda_B) = 2 \delta_{AB}$
and $\lambda_9=\sqrt{2/3}\cdot {\bf 1}$.
Thus, $A=1,\cdots , 8$ 
components are the $SU(3)$ octet, and $A=9$ the singlet. 
However, in the presence of a non-Abelian vortex, it does not 
make sense to mention the multiplets under the $SU(3)_{\rm c+L+R}$
transformation. We rather treat multiplets under the unbroken 
$SU(2)_{\rm c+L+R}$ rotation. The triplet and the singlet are 
respectively given by 
\begin{eqnarray}
\Psi_{\rm t} &\equiv & \Psi^{(1)}\lambda_{1} + \Psi^{(2)}\lambda_{2} 
+ \Psi^{(3)}\lambda_{3}\, ,\label{triplet_def}\\
\Psi_{\rm s} &\equiv & \Psi^{(8)}\lambda_{8} + \Psi^{(9)}\lambda_{9}\, .
\label{singlet_def}
\end{eqnarray}
By using the color-flavor representation, each component 
that appears above is expressed as $\Psi^{(1)}= (d_r+u_g)/\sqrt2$, 
$\Psi^{(2)}= (d_r-u_g)/(\sqrt2 i)$,
$\Psi^{(3)}= (u_r-d_g)/\sqrt2$ for the triplet and 
$\Psi^{(8)}= (u_r+d_g-2s_b)/\sqrt6$, $\Psi^{(9)}=(u_r+d_g+s_b)/\sqrt3 $
for the singlet. 
In addition to the triplet and singlet states, 
there are two `doublet' states. 
However, we do not discuss them here because 
they do not contain zero modes. On the other hand, as we will 
see later, the triplet and singlet states include zero modes 
and are thus important. 

One can explicitly check that $\Psi_{\rm t}$ and $\Psi_{\rm s}$
defined above indeed transform as a triplet and a singlet, 
respectively. 
The $SU(2)_{\rm c+L+R}$ rotation acts on the quark field 
$\Psi$ as 
\begin{eqnarray}
\Psi \rightarrow \Psi'= U_{F}\, \Psi \, U^{\rm T}_{c}
\end{eqnarray}
where $U_{F}={\rm e}^{i \vec{\theta} \cdot \vec{\lambda}/2}$ and 
$U_{c}= {\rm e}^{i \vec{\phi} \cdot \vec{\lambda}/2}$ are the 
$SU(2)_{\rm L/R}$ and $SU(2)_{\rm c}$ rotations, respectively. 
Since we used the vector $\vec{\lambda}=(\lambda_1, \cdots, \lambda_8)$, 
the parameters $\vec{\theta}$ and $\vec{\phi}$ are defined only 
for the first three components 
$\vec{\theta}=(\theta_1, \theta_2, \theta_3, 0, 0, 0, 0, 0)$ and 
$\vec{\phi}=(\phi_1, \phi_2, \phi_3, 0, 0, 0, 0, 0)$. 
For the $SU(2)$ color-flavor locking, the simplest choice is
given by $\phi_{1}=-\theta_{1}$, $\phi_{2}=\theta_{2}$, 
and $\phi_{3}=-\theta_{3}$.
Note also that $\theta_i\ (i=1,2,3)$ are matrices in 
the Nambu-Gor'kov representation 
\begin{eqnarray}
\theta_{1} = \tilde{\theta}_{1}
\left(
\begin{array}{cc}
 1 & 0 \\
 0 & -1
\end{array}
\right), \hspace{1em}
\theta_{2} = \tilde{\theta}_{2}
\left(
\begin{array}{cc}
 1 & 0 \\
 0 & 1
\end{array}
\right),
\hspace{1em}
\theta_{3} = \tilde{\theta}_{3}
\left(
\begin{array}{cc}
 1 & 0 \\
 0 & -1
\end{array}
\right),
\end{eqnarray}
with $\tilde{\theta}_i$ $(i=1,2,3)$ being real numbers.
Then, one finds that the infinitesimal changes of the 
triplet $\Psi_{t}$ components are closed within the three components
$\Psi^{(i)}$ $(i=1,2,3)$:
\begin{eqnarray}
\delta \Psi^{(1)} &=& \theta_{3} \Psi^{(2)} - \theta_{2} \Psi^{(3)}, \nonumber \\
\delta \Psi^{(2)} &=& \theta_{1} \Psi^{(3)} - \theta_{3} \Psi^{(1)},
\label{eq:symmetry_t}
 \\
\delta \Psi^{(3)} &=& \theta_{2} \Psi^{(1)} - \theta_{1} \Psi^{(2)},
\nonumber
\end{eqnarray}
and the singlet $\Psi_{s}$ is invariant
\begin{equation}
\delta \Psi^{(8)} = \delta \Psi^{(9)} = 0\, .
\label{eq:symmetry_s}
\end{equation}

\subsubsection{Zero-mode solutions }

Apart from the complication coming from the color and flavor degrees 
of freedom, the structure of the B-dG equation is essentially the same 
as in the case of the single flavor fermion. Namely, requiring 
finiteness of the solutions with the help of the 
``Majorana-like" condition, one obtains the zero energy solutions 
both in the triplet and singlet states. The B-dG equation naturally 
yields both the left and right modes and the structure of the 
solutions are common in both cases as in the case of the single 
flavor. However, we discuss below only the right-hand modes of 
the zero modes firstly because the chiral symmetry is {\it indirectly} 
broken in the CFL phase (even though it is the $SU(2)$ CFL), 
and secondly because if the system couples to some external 
fields (as the edge states in the quantum hall effect) only one mode (left or right,
depending on the properties of the external field) will 
remain as a zero mode.

The zero modes of the triplet right-handed quarks are 
analytically given by 
\begin{eqnarray}
\Psi^{(1)}(r,\theta)= C_{1}
\left(
\begin{array}{c}
 \varphi(r,\theta) \\
 \eta(r,\theta)
\end{array}
\right), \ 
\Psi^{(2)}(r,\theta) = C_{2}
\left(
\begin{array}{c}
 \varphi(r,\theta) \\
 -\eta(r,\theta)
\end{array}
\right), \ 
\Psi^{(3)}(r,\theta) = C_{3}
\left(
\begin{array}{c}
 \varphi(r,\theta) \\
 \eta(r,\theta)
\end{array}
\right),
\end{eqnarray}
where $C_i$ are normalization constants and 
the particle ($\varphi$) and hole ($\eta$) components are 
\begin{eqnarray}
 \varphi(r,\theta) 
= {\rm e}^{-\int_{0}^{r} |\Delta_{1}(r')|\mbox{d}r'}
\left(
\begin{array}{c}
 J_{0}(\mu r) \\
 i J_{1}(\mu r)\, {\rm e}^{i\theta}
\end{array}
\right), \hspace{0.5em}
 \eta(r,\theta) = {\rm e}^{-\int_{0}^{r} |\Delta_{1}(r')|\mbox{d}r'}
\left(
\begin{array}{c}
 -J_{1}(\mu r)\, {\rm e}^{-i\theta} \\
 i J_{0}(\mu r) \\
\end{array}
\right), \label{tripletZMs}
\end{eqnarray}
for generic shape of the vortex profile $|\Delta_{1}(r)|$. 
Notice that these solutions do not contain the unwinding gap 
$|\Delta_{0}(r)|$. This is easily understood from the structure 
of the B-dG equation (\ref{eq:CFL_eve}). For example, if one looks at 
the $d_r$ and $u_g$ sector, the Hamiltonian contains only $\Delta_1$.
Another peculiarity about this solution is a minus sign of the 
hole component in $\Psi^{(2)}$. However, the minus sign 
must be there so that the zero-mode solutions in the triplet states 
 transform as the adjoint representation under the 
$SU(2)_{\rm c+L+R}$ rotation as shown in Eq.~(\ref{eq:symmetry_t}).

On the other hand, the singlet zero modes
depend on both $|\Delta_{1}(r)|$ and $|\Delta_{0}(r)|$, 
and due to this complication, we did not find explicit 
solutions for generic shape of the gaps. 
Still, approximate solutions are available where 
two gaps $|\Delta_{1}(r)|$ and $|\Delta_{0}(r)|$ 
are both replaced by the constant gap $|\Delta_{\rm CFL}|$ 
in the bulk CFL phase:
\begin{eqnarray}
\Psi^{(8)}(r,\theta) = \frac{D}{\sqrt{6}}
\left(
\begin{array}{c}
 2\varphi_{1}(r,\theta)+\varphi_{2}(r,\theta) \\
 2\eta_{1}(r,\theta)+\eta_{2}(r,\theta)
\end{array}
\right), \hspace{1em}
\Psi^{(9)}(r,\theta) =  \frac{D}{\sqrt{3}}
\left(
\begin{array}{c}
 2\varphi_{1}(r,\theta) - \frac{1}{2}\varphi_{2}(r,\theta) \\
 2\eta_{1}(r,\theta) - \frac{1}{2}\eta_{2}(r,\theta)
\end{array}
\right),
\end{eqnarray}
where $D$ is a normalization constant and  
the particle and hole components are given by
\begin{eqnarray}
 \varphi_{1}(r,\theta) &\simeq& {\rm e}^{-\frac{|\Delta_{\rm CFL}|}{2} r}
\left(
\begin{array}{c}
 J_{0}(\mu r) \\
 i J_{1}(\mu r)\, {\rm e}^{i\theta}
\end{array}
\right), \hspace{0.5em}
 \eta_{1}(r,\theta) \simeq {\rm e}^{-\frac{|\Delta_{\rm CFL}|}{2} r}
\left(
\begin{array}{c}
 -J_{1}(\mu r)\, {\rm e}^{-i\theta} \\
 i J_{0}(\mu r)
\end{array}
\right), \\
 \varphi_{2}(r,\theta) &\simeq& {\rm e}^{-\frac{|\Delta_{\rm CFL}|}{2} r}
\left(
\begin{array}{c}
 J_{0}(\mu r)\, {\rm e}^{-i\theta} \\
 i J_{1}(\mu r) 
\end{array}
\right), \hspace{0.5em}
 \eta_{2}(r,\theta) \simeq {\rm e}^{-\frac{|\Delta_{\rm CFL}|}{2} r}
\left(
\begin{array}{c}
 -J_{1}(\mu r) \\
 i J_{0}(\mu r)\, e^{i\theta}
\end{array}
\right).
\end{eqnarray}
These approximate solutions are in fact asymptotic forms of the 
solutions valid at large distances much away from the vortex 
line.

\begin{figure}[tbp]
\begin{center}
\includegraphics[height=2.2in,keepaspectratio,angle=0]{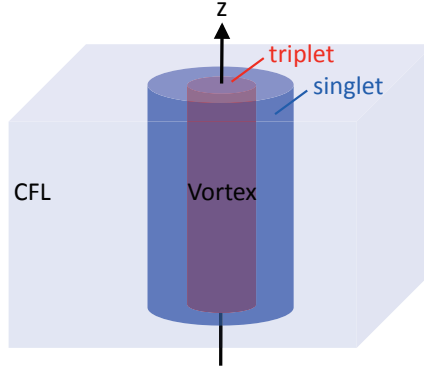}
\end{center}
\caption{Distributions of the triplet and singlet zero modes around
a non-Abelian vortex.}
\label{cylinder}
\end{figure}

In the triplet zero-mode solutions (\ref{tripletZMs}), if one 
approximates the gap profile $|\Delta_1|$ by the constant gap 
$|\Delta_{\rm CFL}|$ in the bulk CFL phase, one finds that the 
solutions decay as ${\rm e}^{-|\Delta_{\rm CFL}|r}$. 
On the other hand, the singlet zero modes decay as 
${\rm e}^{-|\Delta_{\rm CFL}|r/2}$. Therefore, distribution 
of the singlet zero modes is wider than that of the triplet, 
 as schematically shown in Fig.~\ref{cylinder}.

\section{Low Energy Effective Theory of Gapless fermions}
\label{Sect:EFT}

In the previous section, we have discussed strictly zero-energy 
eigenstates of the B-dG equation, and found that such zero modes 
are well-localized on the plane perpendicular to the direction 
of the vortex line. 
While the transverse motion of such zero modes is frozen,
they can move along the vortex line to give a linear dispersion
with respect to the longitudinal momentum $k_z$. 
In this section, we develop an effective description of 
such low-energy excitations. 

\subsection{Single flavor}

We again start with the single flavor case. 
We first notice that the Hamiltonian (\ref{eq:hamil_single}) 
can be decomposed into ``transverse ($\perp$)" and 
``longitudinal ($z$)" parts:
\begin{eqnarray}
{\cal H} &=&
\left(
\begin{array}{cc}
 -i\vec{\alpha}_{\perp} \!\cdot\! \vec{\nabla}_{\perp} - \mu & 
|\Delta| {\rm e}^{i \theta} \gamma_{0} \gamma_{5} \\
 - |\Delta| {\rm e}^{-i \theta} \gamma_{0} \gamma_{5} & -i\vec{\alpha}_{\perp} \!\cdot\! \vec{\nabla}_{\perp} + \mu
\end{array}
\right)
+
\left(
\begin{array}{cc}
 -i \alpha_{z} \frac{\partial}{\partial z} & 0 \\
 0 &  -i \alpha_{z} \frac{\partial}{\partial z}
\end{array}
\right) \nonumber \\
&\equiv& {\cal H}_{\perp} + {\cal H}_{z}.
\end{eqnarray}
Since the longitudinal Hamiltonian ${\cal H}_z$ is linear 
with respect to the $z$ derivative, it immediately implies 
that the low-energy excitations have linear dispersion relations
with respect to $k_z$. Recalling that the zero-mode 
solutions $\Psi_{0}(r,\theta)$ (with $k_z=0$) are in fact 
eigenstates of the transverse Hamiltonian 
$({\cal H}_{\perp}+{\cal H}_z) \Psi_{0}(r,\theta)
={\cal H}_{\perp} \Psi_{0}(r,\theta)=0$, one can see that 
the $z$ dependence enters in a factorized way:
\begin{eqnarray}
\Psi(t,z,r,\theta) = a(t,z) \Psi_{0}(r,\theta)\, ,
\label{long_dep}
\end{eqnarray}
where we have recovered time dependence and consider 
the Schr\"odinger equation $i\partial\Psi/\partial t  ={\cal H}\Psi $. 
Note that the time dependence appears only in $a(t,z)$ because 
the zero-energy states are `static'. 
Plugging Eq.~(\ref{long_dep}) into the Schr\"odinger equation, 
and integrating it over transverse coordinates after multiplying it by
the zero-mode solution $\Psi_0^\dagger$ from the left, we obtain 
the equation of motion for $a(t,z)$: 
\begin{eqnarray}
i \frac{\partial}{\partial t}a(t,z) = \int \Psi_{0}^{\dag}(r,\theta) 
{\cal H}_{z}a(t,z) \Psi_{0}(r,\theta) r\mbox{d}r\mbox{d}\theta,
\label{eq:effective_theory}
\end{eqnarray}
where we have used the normalization 
$\int \Psi_{0}^{\dag} \Psi_{0} r\mbox{d}r\mbox{d}\theta = 1$. 
We can rewrite this equation as 
\begin{eqnarray}
i\left\{ \frac{\partial}{\partial t} + v(\mu,|\Delta|) \frac{\partial}{\partial z} \right\} a(t,z) = 0,
\label{effective_theory}
\end{eqnarray}
with the ``velocity" $v(\mu,|\Delta|)$ defined by 
\begin{eqnarray}
v(\mu,|\Delta|) \equiv \int  
\Psi_0^\dagger(r,\theta) 
\left(
\begin{array}{cc}
 \alpha_{z} & 0 \\
 0 &  \alpha_{z} 
\end{array}
\right) \Psi_0 (r,\theta)\, r\mbox{d}r\mbox{d}\theta \, .
\label{effective_velocity}
\end{eqnarray}
The solution to Eq.~(\ref{effective_theory}) is given by 
$a(t,z)\propto {\rm e}^{i{\cal E}t -ik_z z}$ with 
a linear (gapless) dispersion with respect to $k_z$:
\begin{eqnarray}
{\cal E}=v(\mu,|\Delta|) \, k_{z}.
\end{eqnarray}
Therefore, low-energy excitations inside the vortex 
are gapless (massless) fermions described by 
Eqs.~(\ref{effective_theory}) and (\ref{effective_velocity}). 
One can indeed express these fermions in terms of 
spinors in 1+1 dimensions, and write down an equation 
similar to the Dirac equation.

\begin{figure}[tbp]
\begin{center}
\includegraphics[height=2.2in,keepaspectratio,angle=0]{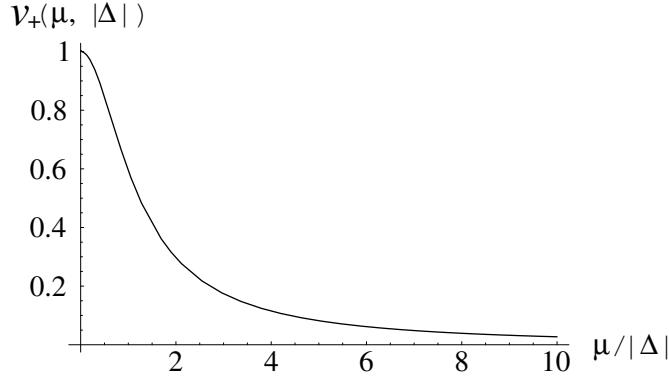}
\end{center}
\caption{Velocity of gapless fermions along the vortex axis.}
\label{velocity}
\end{figure}

One can compute the velocity by using the explicit form of 
the zero-mode solutions, i.e., the right mode 
(Eqs.~(\ref{one_flavor_left_particle}), (\ref{one_flavor_left_hole}))
and the left mode 
(Eqs.~(\ref{one_flavor_right_particle}), (\ref{one_flavor_right_hole})).
If we approximate the gap profile function $|\Delta(r)|$ by 
a constant $|\Delta|$, we obtain for the right mode
\begin{eqnarray}
v_{+}(\mu,|\Delta|) = \frac{\mu^{2}}{|\Delta|^{2}+\mu^{2}} \frac{E(-\frac{\mu^{2}}{|\Delta|^{2}})}{E(-\frac{\mu^{2}}{|\Delta|^{2}})-K(-\frac{\mu^{2}}{|\Delta|^{2}})}-1\, ,
\label{eq:velocity}
\end{eqnarray}
and for the left mode 
\begin{eqnarray}
v_{-}(\mu,|\Delta|) = -v_{+}(\mu,|\Delta|)\, . 
\end{eqnarray}
Here, $K(x)$ and $E(x)$ are the complete elliptic integrals of 
the first and second kinds. 
Therefore, the right (left) mode moves towards the plus (minus) 
direction of the $z$ axis with the velocity less than the speed of light. 
In fact, as shown in Fig.~\ref{velocity}, the velocity $v_{+}(\mu,|\Delta|)$ 
(\ref{eq:velocity}) is a decreasing function of $\mu/|\Delta|$, 
and $v_{+}=1$ (the speed of light) in vacuum ($\mu=0$) [the result 
in vacuum agrees with that obtained by Witten \cite{Witten:1984eb} 
in the context of cosmic strings]. 
This is also seen from the asymptotic forms of the velocity: 
for small $\mu/|\Delta|$
\begin{eqnarray}
v_{+}(\mu,|\Delta|)\simeq 1 -\frac{3}{4}\frac{\mu^2}{|\Delta|^2} 
+ {\cal O}((\mu/|\Delta|)^4),
\end{eqnarray}
while for large $\mu/\Delta$, 
\begin{eqnarray}
v_{+}(\mu,|\Delta|)\simeq \frac{|\Delta|^2}{\mu^2} 
\left(-1 + 2 \ln 2 + \ln \frac{\mu}{|\Delta|}\right)
+ {\cal O}((|\Delta|/\mu)^4)\, . \label{velocity_asymptotic_highmu}
\end{eqnarray}
Since the 
gap in the color superconductivity is parametrically given by 
$|\Delta(\mu)|/\mu \propto {\rm e}^{-b/g(\mu)}$ with $b$ being 
a numerical constant and $g(\mu)$ the running coupling, 
$\mu/|\Delta|$ increases with increasing $\mu$. Therefore, we 
conclude that the effective velocity $v_{+}(\mu)$ decreases with 
increasing $\mu$. If we take typical values $\mu=1000$ MeV and 
$|\Delta|=100$ MeV, then the effective velocity is estimated as 
 $v \simeq 0.027$ which is considerably smaller than the speed of light.

\subsection{Color-Flavor-Locked phase}
\begin{figure}[tbp]
\begin{center}
\includegraphics[height=1.8in,keepaspectratio,angle=0]{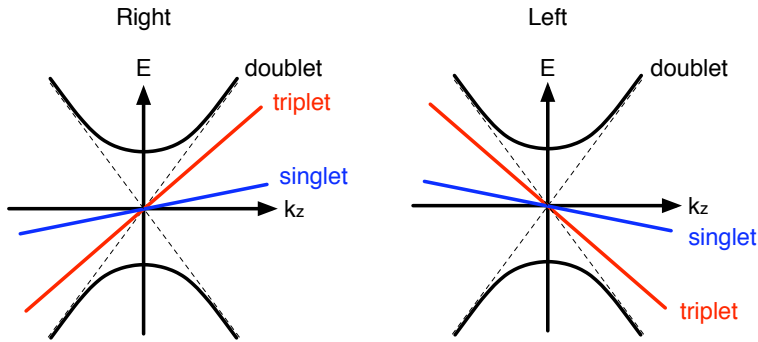}
\end{center}
\caption{Dispersion relations between energy $E$ and momentum in $z$ direction $k_{z}$ for the ground states of triplet, singlet and doublet.}
\label{dispersion}
\end{figure}

For the zero modes trapped in a non-Abelian vortex, one can 
perform the same procedure to obtain the effective theory. 
The low-energy effective theories of the left and right 
fermions in the triplet and singlet states are
\begin{eqnarray}
i\left( \frac{\partial}{\partial t} + v_{i}^{\pm} \frac{\partial}{\partial z} \right) a_{i}(t,z) = 0\, ,
\end{eqnarray}
with the dispersion relations
\begin{eqnarray}
{\cal E}= v_{i}^{\pm} \, k_{z},
\end{eqnarray}
where $i=t,s$ are for triplet and singlet, and $+$, $-$ are for the 
right and left modes. If we approximate the gap 
by a constant value 
in the bulk CFL phase, then the velocity of the triplet is exactly 
the same as the result of the single flavor (Eq.~(\ref{eq:velocity}))
with the replacement of $|\Delta|$ by $|\Delta_{\rm CFL}|$, i.e., 
\begin{eqnarray}
v_{t}^{\pm} =\pm v_{+}(\mu,|\Delta_{\rm CFL}|).
\end{eqnarray}
 On the other hand, since the 
singlet zero modes have a wider transverse dependence, the velocity 
is evaluated by using the same function $v_{+}$ in Eq.~(\ref{eq:velocity}),
but with the replacement of $|\Delta|$ by $|\Delta_{\rm CFL}|/2$, 
i.e., 
\begin{eqnarray}
v_{s}^{\pm} = \pm v_{+}(\mu,|\Delta_{\rm CFL}|/2).
\end{eqnarray} 
In the vacuum $\mu=0$, both the velocities of the 
triplet and the singlet are the same as the speed of light. 
However, with increasing densities, the two velocities start to 
deviate. For example, if one uses the asymptotic 
form of the velocity at large $\mu/|\Delta|$ 
(Eq.~(\ref{velocity_asymptotic_highmu})), then one finds that 
the velocity of the singlet is about quarter of that of the triplet.
Therefore, one can draw a schematic picture of the dispersion 
relations for all the $SU(2)$ multiplets as shown in 
Fig.~\ref{dispersion}.

\section{Summary and discussion}\label{Sect:summary}
We studied the fermion structure of a non-Abelian vortex 
in color superconductivity. By analyzing the Bogoliubov-de Gennes 
equation with the vortex gap profile, we found fermion zero 
modes at the Fermi level in the center of the vortex. We first 
demonstrated the single flavor case with an Abelian vortex 
as a warm-up before the more complicated case, and then 
turned to the CFL case with a non-Abelian vortex. We found 
that there are two gapless modes in the triplet and singlet states
of $SU(2)_{\rm c+L+R} \otimes U(1)_{\rm c+L+R}$
which is the unbroken symmetry in the non-Abelian vortex.
We also constructed the low-energy effective theory of the 
gapless fermions which describes the dynamics of fermions 
living on the vortex line (and thus in 1+1 dimensions).
The gapless modes propagate along the vortex axis at the 
velocities smaller than the speed of light. 

In this paper, we did not consider the presence of any gauge field 
in the vortex. However, the ordinary vortex of type-II 
superconductivity can trap a (quantized) magnetic flux in it. 
In the case of the non-Abelian vortex, there exist gauge fields
which are associated with the unbroken 
$SU(2)_{\rm c+L+R}\otimes U(1)_{\rm c+L+R}$ symmetry. 
However, they are {\it massive} even in the vortex  
because only the winding component of gaps vanishes at 
the center of non-Abelian vortices, 
$\Delta (r=0) = \mbox{diag}(0,*,*)$ with non-zero constants ``$*$", 
where all generators of gauge transformations 
remain broken there. Thus they do not affect the low-energy 
excitations at the energy scale below the masses of gauge bosons. 
Therefore, we expect our effective theory of the 
gapless fermions is valid at least for very low-energy 
excitations. 

In the discussion above, we assumed the limit of massless 
fermions for all flavors. In reality, the strange quark 
has a finite current mass, and cannot be regarded as a 
massless fermion. The effect of the strange quark mass 
on non-Abelian vortices has been studied recently in the 
Ginzburg-Landau model \cite{Eto:2009tr}. In the B-dG equation 
for vortices, we expect the energy-momentum dispersions of 
singlet and triplet zero modes are affected differently 
by this effect since the strange quark differently 
enters them (see Eqs.~(\ref{triplet_def}), (\ref{singlet_def})).

In the Ginzburg-Landau approach, 
a non-Abelian vortex has the ${\bf C}P^2$ {\it bosonic} zero modes 
depending on the flux as explained in Introduction 
\cite{Nakano:2007dr}.
A relation between those bosonic zero modes 
and fermionic zero modes in the B-dG equation 
studied in this paper is unclear at this stage.
We expect that it will be clarified 
if we solve the B-dG equation self-consistently,  
which remains as an important future problem.
It has also been studied \cite{Nakano:2007dr} 
in the Ginzburg-Landau model that 
there exists a repulsive force between 
two parallel non-Abelian vortices, 
and that it does not depend on the ${\bf C}P^2$ zero modes 
if they are placed at a distance much larger 
than their core size. 
This situation is justified for instance in 
a vortex lattice with the lattice spacing larger 
than their core size. 
When a lattice spacing is comparable with the core size, 
we have to use the B-dG equation for multiple vortices. 
We expect that such an analysis gives a force 
depending on the ${\bf C}P^2$ zero modes, namely, on 
color fluxes which are carried by the vortices.

\bigskip
\noindent{\bf Note added:} 
Just before we were about to finish the manuscript, 
we noticed through correspondences with Y. Nishida that he 
was doing very similar calculations as ours. It 
turned out from his recent work \cite{Nishida} that 
his main emphasis was on the topological aspects of the 
fermion zero modes, which we did not discuss in detail. 
Although his analysis is largely the same as ours in the 
single flavor case, he discussed only the Abelian vortices 
in the CFL phase, which is different from our analysis on 
the non-Abelian vortices.

\section*{Acknowledgments}
The authors are grateful to Y. Nishida for correspondences 
and discussions, and to G.E. Volovik for his interest 
in our work and informative comments. One of the authors (KI) 
thanks Y. Kato for explaining him an intuitive picture of zero modes
based on the semi-classical approximation.


\end{document}